\documentclass[fleqn,twoside]{article}
\pdfoutput=1
\usepackage{espcrc2}
\usepackage{epsfig, graphicx, graphics}

\newcommand{\AmS}{{\protect\the\textfont2
  A\kern-.1667em\lower.5ex\hbox{M}\kern-.125emS}}

\title{Brane tilings and supersymmetric gauge theories}

\author{A. Hanany, G. Torri \address[MCSD]{Theoretical Physics Group, The Blackett Laboratory \\
Imperial College London, Prince Consort Road\\ 
London,  SW7 2AZ,  UK \\
\hfill{Imperial/TP/11/AH/02}}}
       
\begin{document}

\maketitle

\section{Supersymmetric gauge theories}

In the past 40 years, gauge theories have played a crucial role in High Energy Physics. In particular, of their biggest successes, has been the formulation of the Standard Model from a gauge principle. 

In this vast sea of theories, the class of supersymmetric gauge theories deserves special attention. In fact, the constraints that supersymmetry imposes, make a gauge theory much more well-behaved and easier to investigate from a theoretical point of view.

Furthermore, supersymmetric gauge theories are shown to arise naturally in Superstring Theory, possibly one of the preferred candidates to solve the problems that were left open by the Standard Model. In particular, the gauge degrees of freedom are brought into the theory by the open strings that are stretched between two D-branes, and are confined precisely on them. Roughly speaking, we could say that, in String Theory, supersymmetric gauge theories arise on the world-volume of D-branes. The interesting thing about this is that the structure of these gauge theories is highly sensitive to the geometry that the D-branes are probing, thus leading to a connection between the shape of space-time and the nature of the interactions of the Universe.

Studying supersymmetric gauge theories in String Theory is not only an interesting endeavour \textit{per se}, but it could also shed some light on the fascinating duality between gauge theories and gravity conjectured by Maldacena more than 10 years ago.

In the following we will be mostly concerned about super-Yang-Mills gauge theories in $(3+1)$ dimensions, although similar considerations can be made for Chern-Simons theories in $(2+1)$ dimensions, \textit{mutatis mutandis}.

\section{Quivers}

As we mentioned above, the structure of supersymmetric gauge theories is highly constrained precisely by supersymmetry. In particular, a supersymmetric lagrangian is unambiguously identified by specifying the number and types of gauge groups, the matter content and a superpotential. 
Furthermore, given the simplicity of the information needed to characterise a supersymmetric gauge theory, it turns out that many of them can be represented by a graph called \textit{quiver}. This is essentially a directed graph containing arrows and nodes with the convention that:
\begin{itemize}
    \item each node represents a gauge group $U(N)$;
		\item each arrow going from a node $a$ to a different node $b$ represents a field $X_{ab}$ in the bifundamental representation $(\mathbf{N},\overline{\mathbf{N}})$ of $U(N_a) \times U(N_b)$.
		\item each loop on a node $a$ represents a field $\phi_a$ in the adjoint representation of $U(N_a)$~.
		\item each superpotential term corresponds to a closed loop in the quiver\footnote{However not all the loops correspond to a superpotential term!};
\end{itemize}

\begin{figure}[htb]
 \includegraphics[height=1.5cm]{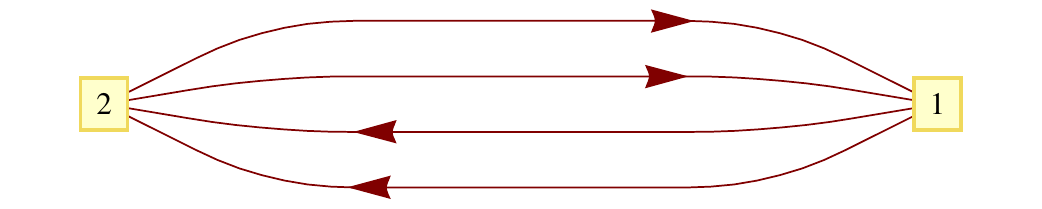}
 \caption{Quiver of the conifold theory}
 \label{fig:quiverconi}
\end{figure}

As an example, Figure \ref{fig:quiverconi} shows the quiver of the well known conifold theory.

The theories that can be represented by a quiver, usually called \textit{quiver gauge theories}, constitute a class of theories which is quite easy to investigate. In fact, they are usually characterised by a very simple superpotential and they contain matter that transform in bifundamental or adjoint representations of the gauge groups.

\section{Brane tilings}
Another way of representing supersymmetric gauge theories is through a type of graph called \textit{brane tiling}. It is a bi-periodic and bi-partite graph with a repeating structure which is called \textit{fundamental domain}. In order to read off from the brane tiling the necessary information to reconstruct the lagrangian, one must keep in mind that:
\begin{itemize}
\item each face corresponds to a $U(N)$ gauge group;
\item each edge corresponds to a bi-fundamental field;
\item each node corresponds to a superpotential term;
\end{itemize}
The nice feature about brane tilings is that they allow one to write down the lagrangian of a supersymmetric gauge theory in a completely unambiguous way\footnote{With a quiver this is not quite true since not all the closed loops correspond to terms in the superpotential.}.

\begin{figure}[htb]
 \includegraphics[height=6cm]{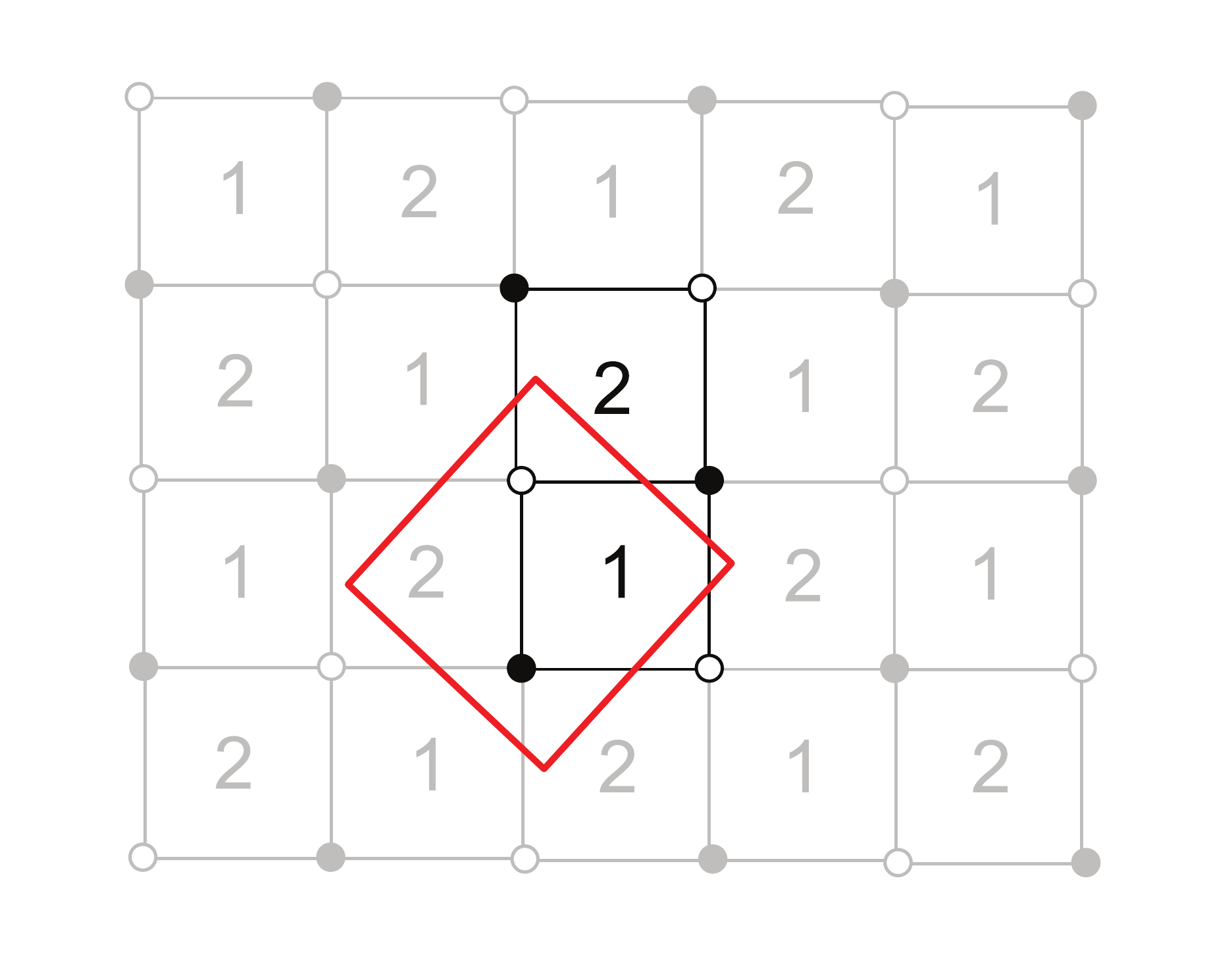}
 \caption{Tiling of the conifold theory}
 \label{fig:tilingconi}
\end{figure}

As an example, Figure \ref{fig:tilingconi} represents the brane tiling of the conifold theory.

Once a supersymmetric gauge theory is specified through a brane tilings, it is possible to investigate relevant features of this theory. In particular, through a technique called \textit{forward algorithm}\footnote{The name suggests the existence of an \textit{inverse algorithm} which, starting from a certain Calabi-Yau, gives the brane tiling of the theory that has that manifold as its vacuum moduli space.}, it is possible to study the vacuum moduli space, i.e. the space of zero-energy solutions to the F-terms and the D-terms of the gauge theory. This algorithm also makes it easy to determine the R-charges of the operators of the theory one is studying.

The brane tilings have been extensively used in the past to investigate a phenomenon called \textit{toric duality} for D3-branes. Roughly speaking, this corresponds to the situation where two (or more) different gauge theories have the same vacuum moduli space. This duality has been shown to be equivalent to Seiberg duality.

Brane tilings seem also like a very natural set-ups to study the \textit{Higgs mechanism} in for supersymmetric gauge theories. In fact, on the tiling this phenomenon is implemented simply as the removal of one or more edges. This makes it very simple also to understand how different gauge theories can be connected through the Higgs mechanism.

Finally, a very fascinating application of brane tilings is the study of supersymmetric Chern-Simons theories in $(2+1)$ dimensions. In 2008, Aharony, Bergman, Jafferis and Maldacena provided the first example of conformal field theory living on an M2-brane probing flat space. With brane tilings it has been possible to extend this duality to more complicated backgrounds and to investigate fascinating features of such theories.

\end{document}